\documentclass[aps,prd,preprint,superscriptaddress,floatfix,preprintnumbers,nofootinbib,longbibliography,a4paper]{revtex4-1}
\pdfoutput=1
\usepackage{color}
\usepackage{graphicx}
\usepackage{textcomp}
\usepackage{subfigure}
\usepackage{feynmp}
\usepackage{amsmath,amssymb,array}
\usepackage{enumerate}
\usepackage{hhline}
\usepackage{soul}
\newcommand{\mathsym}[1]{{}} 
\def\lsim{\:\raisebox{-1.1ex}{$\stackrel{\textstyle<}{\sim}$}\:}
\def\gsim{\:\raisebox{-1.1ex}{$\stackrel{\textstyle>}{\sim}$}\:}

\newcommand{\beqa}{\begin{eqnarray}}
\newcommand{\eeqa}{\end{eqnarray}}
\newcommand{\be}{\begin{equation}}
\newcommand{\ee}{\end{equation}}
\newcommand{\ba}{\begin{array}} 
\newcommand{\ea}{\end{array}}

\begin{document} 
\vspace*{1cm}
 \title{Pseudo-Dirac neutrinos from flavour dependent CP symmetry}
\bigskip
 \author{Anjan S. Joshipura} 
 \email{anjan@prl.res.in}
 \affiliation{Physical Research Laboratory, Navarangpura, Ahmedabad 380 009, India.} 
\author{Ketan M. Patel} 
\email{ketan@iisermohali.ac.in} 
\affiliation{Indian Institute of Science Education and Research Mohali,
Knowledge City, Sector 81, S A S Nagar, Manauli 140 306, India.\\}

\begin{abstract}
\vspace*{0.2cm}
Discrete residual symmetries and flavour dependent CP symmetries consistent
with them have been used to constrain neutrino mixing angles and CP
violating phases. We discuss here role of such CP symmetries in obtaining a
pseudo-Dirac neutrino which can provide a pair of neutrinos responsible
for the solar splitting. It is shown that if  (a) $3\times 3$ Majorana
neutrino matrix $M_\nu$ is invariant under a discrete $Z_2\times Z_2$
symmetry generated by $S_{1,2}$, (b)  CP symmetry $X$ transform $M_\nu$ as
$X^T M_\nu X=M_\nu^*$, and  (c) $X$ and $S_{1,2}$ obey consistency
conditions $X S_{1,2}^* X^\dagger=S_{2,1}$, then two of the neutrino masses are
degenerate independent of specific forms of $X$, $S_1$ and $S_2$. Explicit examples
of this result are discussed in the context of $\Delta(6 n^2)$ groups which
can also be used to constrain neutrino mixing matrix $U$. Degeneracy
in
two of the masses does not allow complete determination of $U$ but
it can also be fixed  once the perturbations are introduced. We consider
explicit
perturbations which break $Z_2\times Z_2$ symmetries but respect CP. These
are shown to remove the degeneracy and provide a predictive description of
neutrino spectrum. In particular, a correlation $\sin
2\theta_{23}\sin\delta_{CP}=\pm {\rm Im}[p]$ is obtained between the atmospheric mixing angle
$\theta_{23}$ and the CP violating phase $\delta_{CP}$ in terms of a group
 theoretically
determined phase factor $p$. Experimentally interesting case
$\theta_{23}=\frac{\pi}{4}$, $\delta_{CP}=\pm \frac{\pi}{2}$ emerges for groups
which predict purely imaginary $p$. We present detailed predictions of the 
allowed ranges of neutrino mixing angles, phases and the lightest neutrino
mass for three of the lowest $\Delta(6 n^2)$ groups with $n=2,4,6$.
\end{abstract}
\maketitle

\section{Introduction}
\label{intro}
Discrete flavour symmetries have been widely used in predicting the observed
patterns of mixing among neutrinos, see
\cite{Altarelli:2010gt,Altarelli:2012ss,Smirnov:2011jv,King:2013eh,
Ishimori:2010au} 
for reviews. Many of the successful schemes
particularly those involving small discrete groups, predict a leading order
mixing patterns which when perturbed can lead to the desired mixing in terms
of very small unknown parameters. Celebrated examples are tri-bimaximal
mixing, bi-maximal mixing or trimaximal mixing among neutrinos which provide
a good description of neutrino mixing at the leading order. For the neutrino
masses, a good leading order description is provided by a pair of degenerate
neutrinos forming the pair responsible for the depletion of the solar
neutrino flux. This leading order mass pattern can also be derived from
discrete flavour symmetries
\cite{Hernandez:2013vya,Joshipura:2014qaa}. See \cite{Wolfenstein:1981kw,Petcov:1982ya,Leung:1983ti} for earlier works on two degenerate neutrinos.

Basic assumptions going in predictions of neutrino mixing are: (i) the
leading order neutrino mass matrix is invariant under a residual $Z_2\times
Z_2$ symmetry (ii) the charged lepton mass matrix $M_lM_l^\dagger$ is
invariant under $Z_m$ with $m>2$ and (iii) both $Z_2\times Z_2$ and $Z_m$ are
embedded in a discrete group $G_f$. The last condition determines the form of the
generators of residual symmetries when leptons are assigned to a definite 3-dimensional irreducible representation of $G_f$. This in turn determines the
exact form of the leptonic mixing matrix $U$ \cite{Lam:2007qc,Lam:2008rs,Lam:2008sh,Lam:2012ga,Lam:2011ag}. One can obtain a pair of
degenerate neutrinos in this set up if the residual symmetry of the neutrino
mass matrix is replaced by $Z_n$, $(n>2)$ group with its generating element
$S$ having the eigenvalues $(\eta,\eta^*,1)$ with $\eta^n=1$ \cite{Joshipura:2014qaa}. This
automatically ensures degeneracy in two of the neutrino masses. Possible
discrete subgroups of $SU(3)$ having such $Z_n$ and the resulting mixing
patterns among neutrinos have been extensively studied in \cite{Joshipura:2014qaa}. We discuss here
an alternative approach in which the occurrence of a pair of degenerate
neutrinos is intimately linked to the imposition of flavour dependent CP
symmetry. Flavour dependent CP symmetries have been widely studied in recent
times  \cite{Feruglio:2012cw,Holthausen:2012dk,Hagedorn:2014wha,Chen:2014wxa,King:2014rwa,Ding:2014ora,Girardi:2015rwa,Li:2016ppt,Yao:2016zev,Joshipura:2018rit} (see \cite{Petcov:2017ggy} for a recent review) with a view to constraint neutrino mixing matrix including Majorana neutrino phases which otherwise remain arbitrary.

In this article, we show that almost the same set up used for predicting
mixing can also be used to obtain either  partially  degenerate
or fully degenerate neutrinos spectrum with a judicious choice of the
flavoured CP symmetry called Generalized CP (GenCP). The analysis presented
here also uses the assumptions (i-iii) outlined above and thus is used also
to constrain leptonic mixing parameters. Exact degeneracy in two of the
masses makes one mixing angle, one Majorana phase and Dirac phase
unphysical. The perturbations introduced to break degeneracy make these
parameters physical and lead to predictions for these quantities which in
general depend on the magnitude and nature of perturbations. We introduce a
class of perturbations which are assumed to be invariant under GenCP. 
This
results into a very predictive framework in which the predicted values of CP
phases are almost independent of perturbations and mostly depend on the
underlying symmetries only. We present some examples of these based on the
$\Delta(6n^2)$ groups. Neutrino mixing patterns resulting from
the assumption of GenCP invariance have been derived earlier in case of the
non-degenerate neutrinos in
\cite{Chen:2014wxa,Chen:2015nha,Li:2017zmk,Joshipura:2018rit}.

We describe in the next section the general conditions under which GenCP symmetry leads to degeneracy in neutrino masses. The specific examples of such symmetries based on $\Delta(6n^2)$ groups are discussed in section \ref{6nsquare}. In section \ref{perturbations}, we discuss a class of perturbations which lead to realistic masses and mixing pattern in a very predictive way and provide a numerical study of such perturbations in section \ref{numerical}. Finally, we summarize in section \ref{summary}.

\section{Degenerate neutrinos from GenCP}
\label{genCP}
We start with the conventional assumptions made in the symmetry based approaches.
\begin{enumerate}[(a)]
\item  Assume that $3\times 3$ Majorana neutrino mass matrix is invariant under
a $Z_2\times Z_2$ symmetry generated by $S_1$, $S_2$:
\be \label{sinv}
S_{1,2}^T\, M_\nu\, S_{1,2}=M_\nu\,.\ee
The above conditions togather imply $S_3^T M_\nu S_3=M_\nu$, where $S_3 =  S_1 S_2$ is also an element of the $Z_2\times Z_2$ group. We also demand that the charged lepton mass matrix $M_l$ is invariant under a $Z_m$ transformation generated by $T_l$ such that
\be \label{tlinv}
T_l^\dagger\, M_lM_l^\dagger\, T_l=M_lM_l^\dagger\,.\ee

\item  $M_\nu$ is invariant under GenCP $X$:
\be \label{xinv}
X^T\, M_\nu\,  X=M_\nu^*\,.\ee

\item CP transformation $X$ followed by a $Z_2\times Z_2$
transformation and an inverse CP transformation is equivalent to a
$Z_2\times Z_2$
transformation on fields and operators. This requires that $X$ maps
$Z_2\times Z_2$ groups to itself
\cite{Feruglio:2012cw,Holthausen:2012dk,King:2014rwa}. This can happen in
three distinct ways. Either
\be \label{xs1s1}
X\, S_i^*\, X^\dagger =S_i\,,\ee
 or 
\be \label{xs1s2}
X\, S_i^*\, X^\dagger =S_j\,,~~X\, S_j^*\, X^\dagger =S_i\,~~{\rm and}~~X\, S_k^*\, X^\dagger =S_k\,,\ee
or
\be \label{xs1s2s3}
X\, S_i^*\, X^\dagger =S_j\,,~~X\, S_j^*\, X^\dagger =S_k\,~~{\rm and}~~X\, S_k^*\, X^\dagger =S_i\,,\ee
where $i,j,k=1,2,3$ and $i\neq j \neq k$. No specific forms for $X$ and $S_i$ are assumed here except that they are unitary and $S_i^2={\bf 1}$, $S_i S_j=S_j S_i$, ${\rm Det.}S_i=1$. The conditions in Eqs. (\ref{xs1s1},\ref{xs1s2},\ref{xs1s2s3}) are special cases of general consistency conditions which are needed to be satisfied for consistent definition of GenCP in a
discrete group $G_f$ \cite{Feruglio:2012cw,Holthausen:2012dk,Ecker:1981wv,Ecker:1983hz,
Bernabeu:1986fc,Chen:2014tpa}:
\be 
X_r\, \rho_r^*(g)\, X_r^\dagger=\rho_r(g')\,,\ee
where $X_r$ is GenCP in representation $\rho_r$ of $G_f$ and $g,g'\in G_f$.
\end{enumerate}
 
We now show that neutrino mass degeneracy directly follows
from these basic assumptions. Specifically, we prove the following result:\\

\emph{If Eqs. (\ref{sinv},\ref{xinv}) are satisfied and $X$ maps $Z_2\times
Z_2$ to itself according to Eq. (\ref{xs1s2})
(Eq. (\ref{xs1s2s3})) then the
absolute masses of
two (all three)  neutrinos are equal.}\\

The proof goes as follows. For simplicity, we consider a specific case
with $i=1$, $j=2$ and $k=3$ in Eq. (\ref{xs1s2}).
Let $U_S$ be a common matrix diagonalizing the commuting matrices 
$S_{1,2}$:
\be \label{d1d2}
U_S^\dagger\, S_{1}\, U_S= {\rm Diag.}(1,-1,-1)\equiv d_1\,,~~U_S^\dagger\, S_{2}\, U_S= {\rm
Diag.}(-1,1,-1)\equiv d_2\,.\ee
The $d_{1,2}$ represent  sets of eigenvalues of $Z_2$ generators $S_{1,2}$.
We have chosen a specific ordering in writing above equations. The other
ordering would give results which can be obtained by permutations of the one
derived here. 
Eqs. (\ref{sinv},\ref{d1d2}) imply
\be\label{mdiag}
d_{1,2}\, D=D\, d_{1,2}\ee
with
\be\label{Ddef}
D=U_S^T\,M_\nu\,U_S\,.\ee
This implies $D={\rm Diag.}(m_1,m_2,m_3).$ The masses $m_i$ are in general complex.
The CP transformation matrix in the diagonal basis of $M_\nu$ becomes
\be \label{deftildex}
\tilde{X} = U_S^\dagger\, X\, U_S^*\,,\ee
and the constraint in Eq. (\ref{xs1s2}) with $i=1$, $j=2$ and $k=3$ then implies
\be \label{xd1d2}
\tilde{X}\, d_{1,2}=d_{2,1}\, \tilde{X}\,.\ee
Using the explicit form of $d_{1,2}$ as given in Eq. (\ref{d1d2}), one finds
that the most general $\tilde{X}$ satisfying Eqs. (\ref{xd1d2}) is
given  by  
\be \label{xtilde}
\tilde{X}=\left(
\ba{ccc}
0&p_1&0\\
p_2&0&0\\
0&0&p_3\\ \ea \right)\,.\ee
where $p_i$ are arbitrary phases.
Eq. (\ref{xinv}) can be rewritten using Eqs. (\ref{Ddef},\ref{deftildex}) as
\be \label{keyeq}
\tilde{X}^T\, D\, \tilde{X}=D^*\,.\ee
Eqs. (\ref{xtilde},\ref{keyeq}) give:
\be \label{ratio}
\frac{m_2^*}{m_1}=p_1^2=p_2^2\,,~\frac{m_3^*}{m_3}=p_3^2\,.\ee
Since $p_{1,2}$ are phases, Eq. (\ref{ratio}) immediately leads to 
degeneracy of two masses
$$|m_1|=|m_{2}|$$
proving the above assertion. 

Following the similar reasoning, it is easy to verify that a choice
$i=2$, $j=3$, $k=1$ in Eq. (\ref{xs1s2}) leads to $|m_2|=|m_3|$ while $i=3$,
$j=1$, $k=2$ in the same equation would imply $|m_1|=|m_3|$. On the other hand, if Eq. (\ref{xs1s2s3}) is true then the same reasoning as
above leads to
\be \label{xtilde2}
\tilde{X}=\left(
\ba{ccc}
0&0&p_1\\
p_2&0&0\\
0&p_3&0\\ \ea \right)\,~~{\rm or}~~\tilde{X}=\left(
\ba{ccc}
0&p_1&0\\
0&0&p_2\\
p_3& 0 &0\\ \ea \right)\,.\ee
Substitution of either of the above $\tilde{X}$ in Eq. (\ref{keyeq}) leads to equality of the absolute
masses of all three neutrinos. In the following, we confine ourselves to
discussion of partially degenerate spectrum.

Apart from implying degeneracy, Eq. (\ref{keyeq}) also fixes phases of two of the masses in terms of $p_{1,2,3}$. The latter are determined by the structure of $X$ and would follow once the forms of $S_{1,2}$ and the underlying discrete groups are fixed. Explicitly, let $m_1=m e^{i\alpha}$ with $m$ real and positive. Then 
\be \label{realmass}
m_2=m e^{-i\alpha}p_1^{-2}\,,~ m_3=m' p_3^{-1}~.\ee
Defining
\be \label{pnu}
P_\nu={\rm Diag.} (  e^{-\frac{i\alpha}{2}}, 
e^{\frac{i\alpha}{2}} p_1, p_3^{\frac{1}{2}})\,,\ee
$U_R=U_S P_\nu$  makes the neutrino mass matrix diagonal with real positive\footnote{We are assuming $m'$ to be positive. If not, it can be made positive by redefining the $(3,3)$ element of $P_\nu$.} entries.
Now using Eq. (\ref{Ddef}) one gets
\be \label{dr}
U_R^T\, M_\nu\, U_R\equiv {\rm Diag.} (m,m,m')\,.\ee
In this basis, $\tilde{X}_R \equiv U_R^\dagger X U_R^*$ has the form
\be \label{xtilder}
\tilde{X}_R=\left( \ba{ccc}
0&1&0\\
(-1)^l&0&0\\
0&0&1\\ \ea \right)\,,\ee
with $l=0,1$ as can be explicitly checked using
Eqs. (\ref{xtilde},\ref{pnu}). It is
remarkable that there are only two possible forms of $\tilde{X}_R$
irrespective of the underlying discrete group or the chosen residual
symmetry. This result follows from the required consistency equations in a
model independent manner. The other possible $\tilde{X}_R$ are either
related to the 
above by permutations or by an overall change of sign. $\tilde{X}_R$ with
$l=0$ is symmetric and hence the corresponding $X$ is symmetric. On the
other hand $X$ in case of $l=1$ is neither symmetric or antisymmetric.

Conversely, if we assume two of the neutrinos to be degenerate then one
could argue that the allowed forms of $\tilde{X}_R$ are still given by Eq.
(\ref{xtilder}). Going to the basis with real and positive masses, Eq.
(\ref{keyeq}) becomes
$$\tilde{X}_RD_R=D_R \tilde{X}_R\,,$$ 
where $D_R$ is a diagonal matrix with real and positive entries given by $|m_i|$. If we assume
$|m_1|=|m_2|$ the  allowed forms of $\tilde{X}_R$ satisfying above equations  are given by
\be \label{xrth}
\tilde{X}_R=\left(\ba{ccc}
\cos \theta & \sin\theta&0\\
-\sin\theta & \cos\theta&0\\
0&0& \pm1\\ \ea \right)\,;~~\tilde{X}_R=\left(\ba{ccc}
\cos \theta & \sin\theta&0\\
\sin\theta & -\cos\theta&0\\
0&0& \pm1\\ \ea \right)\,.\ee
The above form of $\tilde{X}_R$ implies existence of an unbroken $O(2)$
symmetry in the neutrino mass basis. The consistency conditions 
$\tilde{X}_R\, d_{1,2}=d_{2,1}\, \tilde{X}_R$ which follow from Eq.
(\ref{xd1d2}) and definition of $\tilde{X}_R$ lead to $\theta=\frac{\pi}{2}$
in both the cases and the allowed forms of $\tilde{X}_R$ reduce to Eq.
(\ref{xtilder}). The consistency conditions break the full $O(2)$ symmetry
into its discrete subgroups, $Z_2$ and $Z_4$, generated by $\tilde{X}_R$ in
Eq. (\ref{xtilder}) for $l=0$ and $l=1$ respectively. We note that:
\begin{itemize}
\item Degeneracy here follows from Eq. (\ref{xs1s2}).
If $Z_2\times Z_2$ is mapped by $X$ instead according to Eq. (\ref{xs1s1})
then it is easy to show that the corresponding $\tilde{X}_R$ is proportional
to an identity matrix instead of non-diagonal $\tilde{X}_R$ as in Eq.
(\ref{xtilder}). In that case,  Eq. (\ref{xinv})
does not imply any restriction on the neutrino masses.
\item A class of discrete symmetric groups leading to a degenerate mass
spectrum were identified and studied in \cite{Joshipura:2014qaa} by
requiring that these groups  posses a generator $S$ with eigenvalues
$(\eta,\eta^*,1)$, where $\eta^m=1$ and $m>2$. The chosen form of $S$ 
corresponds to non-trivial discrete subgroups of $O(2)$ which ensures
degeneracy in the masses of the first two generations of neutrinos. The
mechanism for obtaining degenerate spectrum  through CP symmetry presented
here appears logically different but invariance under $S$ would nevertheless
be present since any neutrino mass matrix with two degenerate eigenvalues is
always invariant in a suitable basis under a  symmetry generated by $S$.
This symmetry thus is an effective symmetry in this approach.
\end{itemize}

In the next section, we discuss specific groups and CP symmetries as concrete realization of above general discussions.

\section{Degenerate pair of neutrinos in $\Delta(6 n^2)$ groups}
\label{6nsquare}
The $\Delta(6 n^2)$ as possible flavour symmetry
groups are widely studied
\cite{Escobar:2008vc,King:2013vna,Joshipura:2016quv}. In particular, all
possible CP  and residual neutrino  $Z_2\times Z_2$ symmetries in $\Delta(6
n^2)$ have been extensively studied with a view to constrain neutrino
Majorana phases \cite{King:2014rwa}. Here we show that a subclass of CP
symmetries identified in \cite{King:2014rwa} actually satisfy the conditions
of theorem derived in the previous section and hence lead to a degenerate
pair of neutrinos. For this purpose, we closely follow the notation of
\cite{King:2013vna,King:2014rwa}.

The groups in $\Delta(6 n^2)$ series can be constructed from three generators $a,b,c$ satisfying 
$$a^3=b^2=(ab)^2=c^n=1~.$$ 
It is convenient to define $d=a^2c a$ satisfying $a c a^2=c^{-1}d^{-1}$. We shall be using a specific three dimensional representation for these:
\be \label{represent}
a=\left(\ba{ccc}
0&1&0\\
0&0&1\\
1&0&0\\ \ea\right),~b = -\left(\ba{ccc}
0&0&1\\
0&1&0\\
1&0&0\\ \ea\right),~c = \left(\ba{ccc}
\eta & 0 &0\\
0& \eta^* &0\\
0& 0 &1\\ \ea\right),~d= \left(\ba{ccc}
1 & 0 &0\\
0& \eta &0\\
0& 0 & \eta^*\\ \ea\right), \ee
with $\eta^n=1$. We consider specific Klein groups $K=(1,\, c^{\frac{n}{2}},\, abc^\gamma,\,
abc^{\gamma+\frac{n}{2}})$ with $\gamma=0,1,...,\frac{n}{2}-1$ which are
obtained as subgroups of an underlying $\Delta(6 n^2)$ group with only even
values for $n$. Of these, we specifically choose the neutrino
symmetries as
\be \label{s1s2}
S_1=abc^\gamma=\left(\ba{ccc}
0&-\eta^{-\gamma}&0\\
-\eta^\gamma &0 & 0\\
0&0&-1\\\ea\right);~~ S_2=abc^{\gamma+\frac{n}{2}}=\left(\ba{ccc}
0 &\eta^{-\gamma} &0\\
\eta^\gamma & 0 & 0\\
0&0&-1\\\ea\right)~.\ee
The allowed set of CP symmetries satisfying the consistency conditions are also identified in
\cite{King:2014rwa} as 
\be \label{cp}
(X_1,X_2,X_3,X_4)\equiv (c^x d^{2\gamma+2x},\, a bc^xd^{2 x},\, c^x d^{2 x+2\gamma+\frac{n}{2}},\, ab c^x d^{2 x+2\gamma+\frac{n}{2}})~,\ee
where $x=0,1,...,n-1$ and we have neglected an overall phase. All the four $X_i$ map the Klein group into itself and $X_{3,4}$ satisfy Eq. (\ref{xs1s2}):
$$ X_{3,4}\, S_{1,2}^*\,  X^\dagger_{3,4} =S_{2,1}\,.$$
This can be verified from the explicit 3-dimensional representation in Eq. (\ref{s1s2}) and the form of $X_{3,4}:$ 
\be \label{x3x4}
X_3=\left(\ba{ccc}
\eta^x & 0 & 0\\
0 & - \eta^{x + 2 \gamma} &0\\
0 & 0 & -\eta^{-2 x - 2 \gamma} \\ \ea\right)\,;~~X_4=\left(\ba{ccc}
0&\eta^{x}&0\\
-\eta^x&0&0\\
0&0&\eta^{-2 x}\\ \ea\right)\,.\ee
Both these symmetries must lead to a  degenerate pair. This is explicit from the construction of neutrino mass matrices satisfying Eqs. (\ref{sinv},\ref{xinv}) with $X$ as $X_3$ and $X_4$. 
In case of $X_3$ one gets 
\be \label{mnu3}
M_{0\nu}=m_0\left(
\ba{ccc}
A \eta^{-x}& i B \eta^{-x -\gamma}&0\\
iB \eta^{-x-\gamma}&A \eta^{-x -2 \gamma}&0\\
0&0&C \eta^{2 x+2 \gamma}\\
\ea\right)~.\ee
Here $A$, $B$, $C$ are required to be real.
For $X_4$, imposition of Eq. (\ref{xinv}) leads to the following form:
\be \label{mnu4}
M_{0\nu}=m_0\left(
\ba{ccc}
A \eta^{-x}&  i B \eta^{-x }&0\\
i B \eta^{-x}& A \eta^{-x }&0\\
0&0&C \eta^{2 x}\\
\ea\right)~,\ee
with real $A,B,C$. These matrices lead to a degenerate
pair\footnote{Neutrino mass matrices invariant under $X_{3,4}$ and $S_{1,2}$
were earlier derived in \cite{King:2014rwa}. These matrices did not contain
a relative factor of $i$ present here between the off-diagonal and diagonal elements in Eqs. (\ref{mnu3},\ref{mnu4}).
This factor leads to degeneracy which should occur in these cases as argued
on general grounds.}. It is to be noted that just the generalized CP
invariance alone is sufficient to give degeneracy in case of $X_4$.
Imposition of the residual symmetries $S_{1,2}$ invariance further leads to
$A=0$ in Eq. (\ref{mnu4}) for $\gamma\not=0$ in $S_{1,2}$. For $\gamma=0$ in $S_{1,2}$, Eq.
(\ref{mnu4}) is also invariant under $S_{1,2}$.

\section{GenCP invariant perturbations}
\label{perturbations}
The residual  Klein symmetry along with particular $Z_m$ symmetry of
$M_lM_l^\dagger$ lead to prediction of vanishing Dirac CP phase
\cite{King:2013vna,Joshipura:2016quv} in $\Delta(6 n^2)$ groups.  The
additional use of GenCP further lead to predictions of the Majorana phases
\cite{King:2014rwa}. These predictions hold strictly only for the
non-degenerate neutrino masses and would not be true in cases of  specific
CP symmetries $X_{3,4}$ which imply degeneracy since  in this case Dirac CP
and one of the Majorana CP phases can be rotated away. One therefore
needs perturbations which break degeneracy. We consider here a very specific
class of perturbations which break the residual $Z_2\times Z_2$ symmetry of
$M_\nu$ but preserve the GenCP invariance. As we will show, this class of
perturbations lead to a departure from degeneracy but  lead to very definite
prediction of Dirac CP phase in terms of the atmospheric mixing angle
irrespective of the values of perturbing parameters.  

We assume the  neutrino mass matrix $M_\nu$ to be
\be \label{mnu}
M_\nu=M_{0 \nu}+\delta M_\nu~.\ee
Here, $M_{0\nu}$ is assumed to satisfy Eqs. (\ref{sinv},\ref{xinv}) and thus has the form given in Eq. (\ref{mnu3}). $\delta M_\nu$ only satisfies Eq. (\ref{xinv}):
\be \label{delmnu}
X_3^T\, \delta M_\nu\, X_3=\delta M_\nu^*~.\ee   
We do not consider here $X_4$ since in this case the  most general matrix
$M_\nu$ invariant under $X_4$ alone leads to degeneracy as shown in the
earlier section. Types of perturbations considered here are thus not
sufficient to give correct neutrino mass spectrum in case of $X_4$.

The explicit form of $X_3$ given in Eq. (\ref{x3x4}) leads to the following general
neutrino mass matrix:
\be \label{explicitmnu}
M_{\nu}=m_0\left(
\ba{ccc}
A \eta^{-x}& i B \eta^{-x -\gamma}&i \epsilon_1\eta^{\gamma+\frac{x}{2}}\\
iB \eta^{-x-\gamma}&A(1+\epsilon_3) \eta^{-x -2 \gamma}& \epsilon_2\eta^{\frac{x}{2}}\\
i \epsilon_1\eta^{\gamma+\frac{x}{2}}&\epsilon_2\eta^{\frac{x}{2}}&C \eta^{2 x+2 \gamma}\\
\ea\right)~.\ee
Here $\epsilon_{1,2,3}$ are required to be real. We have also redefined
$A,B,C$ to absorb some of the redundant parameters implied by perturbations.
As seen from above, the general perturbations in this case are
characterized by three real parameters. It was shown in
\cite{Li:2017zmk,Joshipura:2018rit} that GenCP invariant perturbations in
case of
$X_{1,2}$ can be described by an orthogonal matrix containing three angles.
This is true in this case also. We notice that $M_\nu$ in Eq.
(\ref{explicitmnu}) can be made real by multiplying it with a diagonal phase
matrix $P:$
\be \label{pnu1}
P={\rm Diag.} (\eta^{\frac{x}{2}},-i
\eta^{\frac{x}{2}+\gamma},i\eta^{-x-\gamma})~.\ee
\be \label{mnureal}
P^T M_\nu P=m_0\left(
\ba{ccc}
A &  B&-\epsilon_1\\
B&- A(1+\epsilon_3)& \epsilon_2\\
 -\epsilon_1&\epsilon_2& - C\\
\ea\right)~.\ee
This matrix being real and symmetric, can be diagonalized by an orthogonal matrix. It follows from Eqs. (\ref{explicitmnu},\ref{mnureal}) that the matrix diagonalizing $M_\nu$ has the form
\be \label{unu}
U_\nu= P OK~,\ee
where $O$ is a real orthogonal matrix which diagonalizes the matrix, Eq (\ref{mnureal}) and $K$ is a diagonal matrix with elements $\pm 1$ or $\pm i$ which is introduced to make the eigenvalues of $M_\nu$ positive. The leptonic mixing matrix is given by
\be \label{upmns}
U_{\rm PMNS}\equiv U =U_l^\dagger U_\nu=U_l^\dagger P OK~.\ee
Here $U_l$ can be determined from Eq. (\ref{tlinv}) in terms of the matrix which diagonalize $T_l$.   A diagonal $T_l$ implies a trivial $U_l$ which leads to vanishing $\theta_{23}$ and $\theta_{13}$ in the absence of perturbations as can be seen from Eqs. (\ref{mnu3},\ref{unu},\ref{upmns}). Viable $\theta_{23}$ would then require large corrections and therefore we do not consider this case. There exist two classes of non-diagonal $Z_m$ in $\Delta(6 n^2)$ which can be chosen as $T_l$. These have the general form in the basis defined by Eq. (\ref{represent}):
\be
T_l= \left(\ba{ccc}
0&\eta_1&0\\
0&0&\eta_2\\
\eta_3&0&0\\
\ea\right) ~~{\rm or}~~ T_l=-\left(\ba{ccc}
0&0&\eta_1\\
0&\eta_3&0\\
\eta_2&0&0\\
\ea\right) ~,\ee
with $\eta_1\eta_2\eta_3=1.$
The first set of $T_l$ is chosen as a proper symmetry in the discussion of
non-degenerate neutrinos \cite{King:2013vna,Joshipura:2016quv,King:2014rwa}.
Here we argue that the second type of $T_l$ or its permutations  are the
only viable choices if we assume that perturbations are small and their role
is to provide small corrections to zeroeth order mixing. For this, we note
that $O$ is a block diagonal matrix with $O_{a3},O_{3a}$ zero for $a=1,2$ in
the absence of perturbations, see the form of $M_{0\nu}$, Eq. (\ref{mnu3}).
The third column of $O$ is thus given by $\psi=(0,0,1)^T$ and this represents
the eigenvector for the non-degenerate mass. It
then follows that $U_l^\dagger \psi$ should provide a good leading order
approximation to the third column of $U$. If $T_l$ is chosen as the first
set of matrix then the absolute value of all the elements of $U_l$ is
$\frac{1}{\sqrt{3}}$ and the $|U_l^\dagger \psi|$ is also democratic
implying $\sin^2\theta_{13}=\frac{1}{3}$. This would require large corrections. If $T_l$ is chosen in the second 
category, then the corresponding $U_l$ can be written as 
 \be \label{ul}
U_l=\frac{1}{\sqrt{2}}\left(\ba{ccc}
0&1&-\eta_1 \lambda^*\\
\sqrt{2}&0&0\\
0&\lambda\eta_1^*&1\\
\ea\right)~.\ee
Here, $\lambda=(\eta_1\eta_2)^{\frac{1}{2}}$ and three columns of above $U_l$ correspond to eigenvectors with eigenvalues $-\lambda^{*2}, \lambda, -\lambda$ respectively. Given this $U_l$,
$|U_l^\dagger \psi|$ is given by $\frac{1}{\sqrt{2}} (0,1,1)$ corresponding
to vanishing (maximal) $\theta_{13}$ $(\theta_{23})$ at the leading order.
This can get corrected to the desired values after small  perturbations.

If neutrinos are non-degenerate, then above $U_l$ alongwith $S_{1,2}$
given in Eq. (\ref{s1s2}) leads to bi-maximal mixing pattern as long as $T_l$
is not $Z_2$. This predicts maximal solar angle which would need to be
corrected by large perturbations. In the present case, solar angle remains
undetermined at the leading order due to degeneracy and could get corrected
by small perturbations.

One can construct the mixing matrix $U$ using $U_\nu$ and $U_l$ respectively
from Eqs. (\ref{unu}) and  (\ref{ul}). The elements of $U$ are given as
follows:
\beqa \label{uelements}
U_{ej}&=&-i \eta^{\gamma+\frac{x}{2}}O_{2j}\,,\nonumber\\
U_{\mu j}&=&\frac{\eta^{\frac{x}{2}}}{\sqrt{2}}(O_{1j}+p\, O_{3j})\,,\nonumber\\
U_{\tau j}&=&\frac{-\lambda \eta_1^*\eta^{\frac{x}{2}}}{\sqrt{2}}(O_{1j}-p\, O_{3j})\,,\eeqa
where $j=1,2,3$ and $p=i \lambda^*\eta_1\eta^{-\gamma-\frac{3x}{2}}$. By
comparing above $U$ with standard parameterization of it as given by
\cite{Patrignani:2016xqp} one arrives at the following
relations:
\beqa \label{correlations}
\sin2\theta_{23}\sin\delta_{CP}&=&\pm {\rm Im}[p]\,,\nonumber\\
\sin\alpha_{21}&=&\sin(\alpha_{31}-2\delta_{CP})=0~,\nonumber\\
\cos2\theta_{23}&=&\sin\chi\, \left(1-{\rm Im}[p]^2\right)^{\frac{1}{2}}~,\eeqa
where $\tan\chi=\frac{O_{13}}{O_{33}}$. It is seen from the first two 
equations that the  CP violating phases are predicted solely in terms of a
group theoretical phase $p$ and $\theta_{23}$. The latter depends upon
perturbation if $p$ is not purely imaginary. One obtains the $\mu$-$\tau$
reflection symmetry \cite{Grimus:2003yn,Kitabayashi:2005fc,Farzan:2006vj,Joshipura:2009tg,Gupta:2011ct,Joshipura:2015dsa,He:2015xha,Chen:2015siy} (see also \cite{Xing:2015fdg} for a recent review)
results 
$$\pm \delta_{CP}=2\theta_{23}=\frac{\pi}{2}\,,~\sin\alpha_{21}=\sin\alpha_{31}=0$$
for special class of residual symmetries for which $p$ is purely imaginary. While this symmetry is close to the observed values there already exists hints on its small breaking since the best fit values of $\theta_{23}$ is found to be different from the maximal values for both the orderings in neutrino masses \cite{Esteban:2016qun}. 
\begin{figure}[!ht]
\centering
\subfigure{\includegraphics[width=0.50\textwidth]{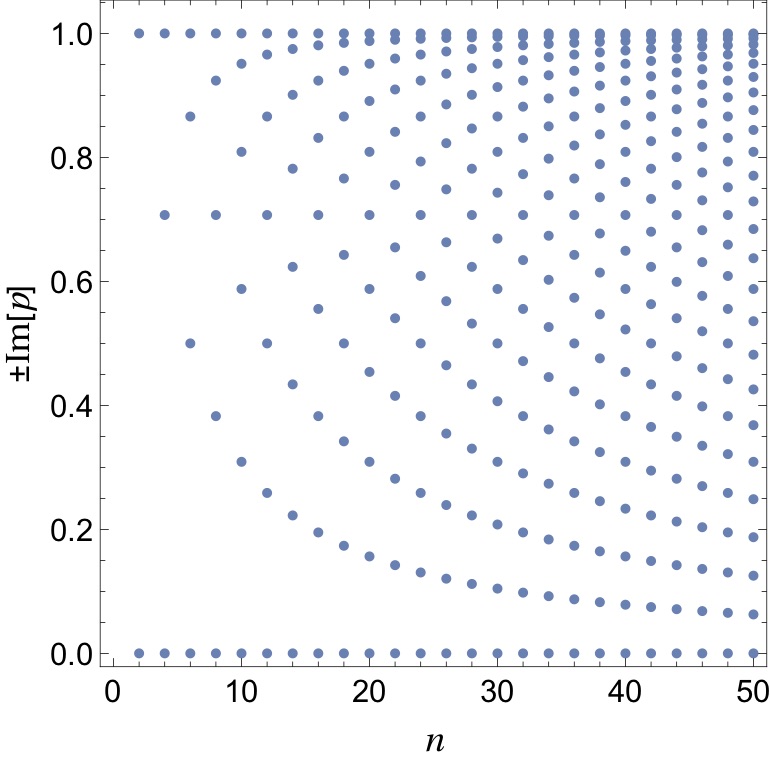}}
\caption{The values of $\pm {\rm Im}[p]$ predicted by the groups of $\Delta(6n^2)$ series for even $n$, and $n \le 50$.}
\label{fig1}
\end{figure}
For $n=2$, the group $\Delta(6\times 2^2) \equiv S_4$ leads to either real
or pure imaginary $p$ as can be seen from Fig. \ref{fig1}. The first of
these implies $\delta_{\rm CP} = 0$ while the latter leads to residual
unbroken $\mu$-$\tau$ reflection symmetry even in the presence of
perturbations as discussed above. Groups with higher $n$ contain
other values of $p$ in additions to the ones in $S_4$. For example, $n=4$
has ${\rm Im}[p] = \pm \frac{1}{\sqrt{2}}$ and $n=6$ contains ${\rm Im}[p] = \pm \frac{1}{2}$ and $\pm \frac{\sqrt{3}}{2}$.
These values may be taken as approximate predictions of $\sin\delta_{CP}$ since the allowed 3$\sigma$ range for $\sin 2 \theta_{23}$ is quite narrow ($0.97$-$1.0$) \cite{Esteban:2016qun} and hence  $\sin\delta_{CP}$ is nearly close to ${\rm Im}[p]$. The exact values depend on details of perturbations. We study these in the next section.

We note that correlation similar to the first of
Eq. (\ref{correlations}) was also derived in \cite{Chen:2015siy}  by imposing a specific CP symmetry
on the neutrino mass matrix in the 
flavour basis. Imposing our CP symmetry $X_3$ in symmetry basis is
equivalent to imposing
a symmetry $X_{3f}\equiv U_l^\dagger X_3U_l^*$ in the flavour basis with
diagonal charged lepton mass matrix, where $U_l$ is given by Eq. (\ref{ul}).
Explicitly
\begin{equation}\label{x3f}
X_{3f}=\left(\begin{array}{ccc}
\eta^{x+2\gamma}&0&0\\[.25cm]
0&e^{i\beta} \cos \theta&i e^{i\frac{\beta+\alpha}{2}} \sin \theta\\[.25cm]
0&i e^{i\frac{\beta+\alpha}{2}} \sin \theta&e^{i\alpha} \cos \theta\\[.25cm]
\end{array}\right)~,\end{equation}
where $\sin\theta={\rm Im}[p]={\rm Im}[i
\lambda^*\eta_1\eta^{-\gamma-\frac{3x}{2}}]$, $e^{i\beta}=\lambda^*\eta_1\eta^{
-x/2-\gamma}$ and $e^{i\alpha}=\lambda\eta_1^*\eta^{-x/2-\gamma}$. This has the
same form as the generalized $\mu$-$\tau$ reflection symmetry introduced in
\cite{Chen:2015siy} and correlation derived by them coincide with our
Eq. (\ref{correlations}) with an important difference.
$\theta$, $\beta$, $\gamma$ introduced by them are arbitrary parameters while here
they are determined
by group theory. As a result, one  predicts definite pattern of the
$\mu$-$\tau$ symmetry  breaking since the phase factor $p$ takes specific
discrete values based on the chosen group and residual symmetries. 
The values of real and imaginary parts of $p$ predicted by the first few groups of the group series $\Delta(6 n^2)$ which possess 3-dimensional irreducible representations are displayed in Fig. \ref{fig1}. We have displayed only the groups with even $n$ since the groups with odd $n$ do not contain any Klein groups as subgroups. 

\section{Numerical study of GenCP invariant perturbations}
\label{numerical}
\begin{figure}[!ht]
\centering
\subfigure{\includegraphics[width=0.48\textwidth]{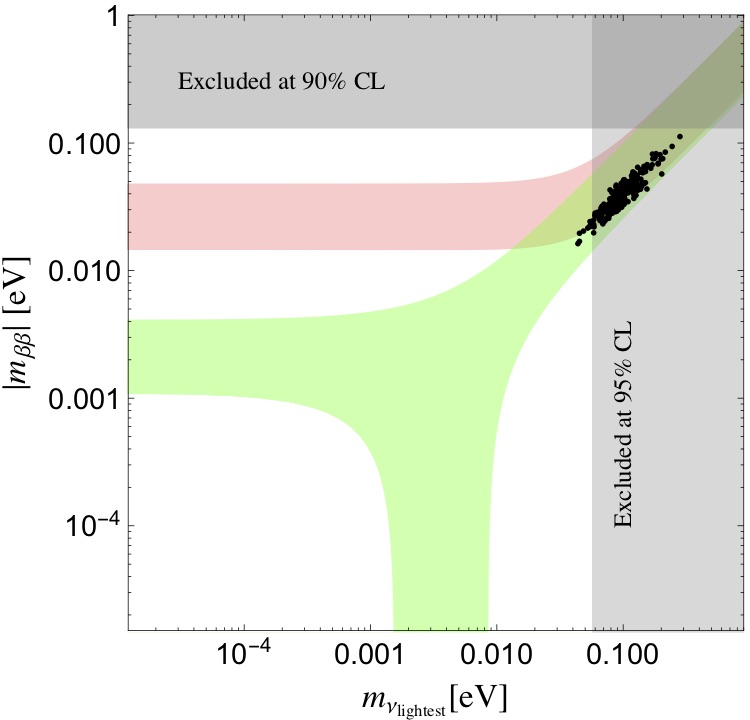}}\hspace*{0.1cm}
\subfigure{\includegraphics[width=0.48\textwidth]{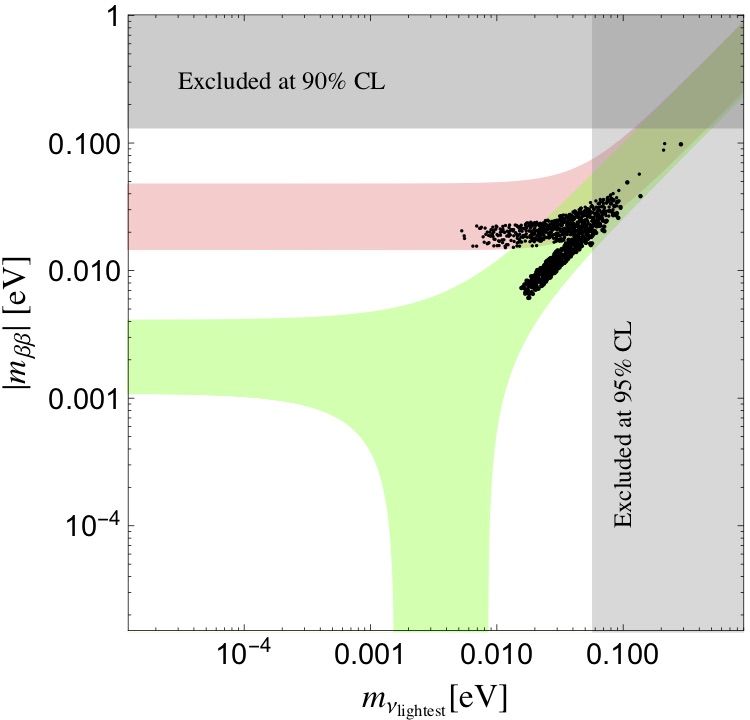}}
\caption{Correlations between the effective neutrinoless double beta decay mass and the lightest neutrino mass in case of $n=2$ (with $\gamma=0$, $x= - \eta_1 = \eta_2 = 1$). The left (right) panel corresponds to $\epsilon_{\rm max} = 0.02$ ($0.1$). The black points in both the panels are in agreement with the results of a global fit \cite{Esteban:2016qun} of neutrino oscillation parameters within $3\sigma$.}
\label{fig2}
\end{figure}

\begin{figure}[!ht]
\centering
\subfigure{\includegraphics[width=0.48\textwidth]{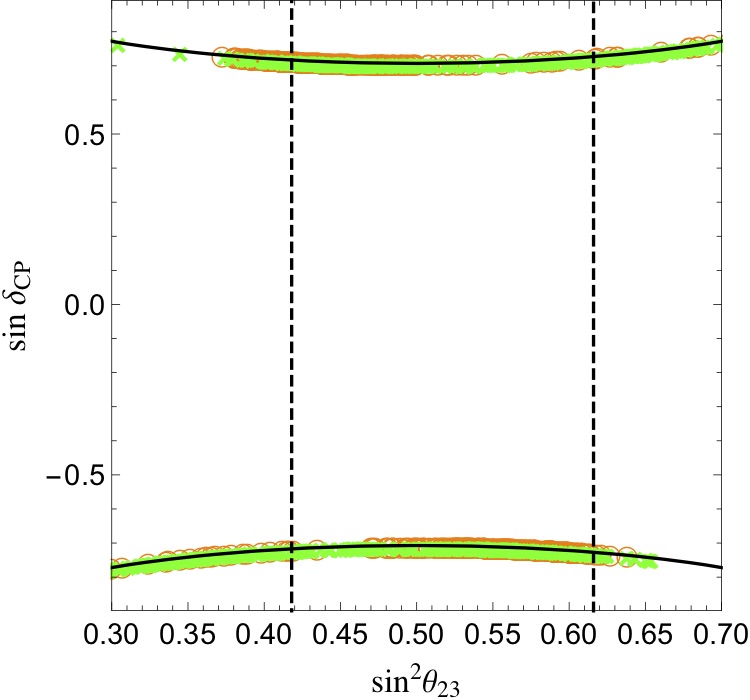}}\hspace*{0.1cm}
\subfigure{\includegraphics[width=0.48\textwidth]{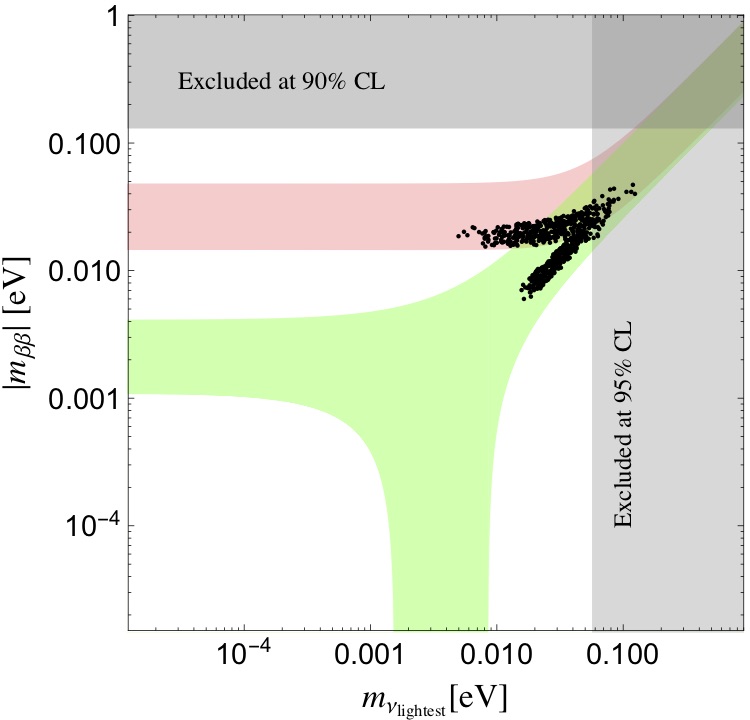}}
\caption{The results for $n=4$ (with $\gamma=0$, $x= - \eta_1 = \eta_2 = 1$) and $\epsilon_{\rm max}=0.1$. In the left panel, the orange circles (green crosses) correspond to the normal (inverted) ordering in the neutrino masses. All these points are in agreement with global fit values of $\Delta m^2_{21}$, $\Delta m^2_{31}$, $\theta_{12}$ and $\theta_{13}$ within $3\sigma$. The region between the vertical dashed lines correspond to experimentally allowed $3\sigma$ range of $\sin^2\theta_{23}$. The black contours represent the correlation given by the first of Eq. (\ref{correlations}). In the right panel, all the points are in agreement with the results of global fit of neutrino oscillation data within $3\sigma$.}
\label{fig3}
\end{figure}

\begin{figure}[!ht]
\centering
\subfigure{\includegraphics[width=0.48\textwidth]{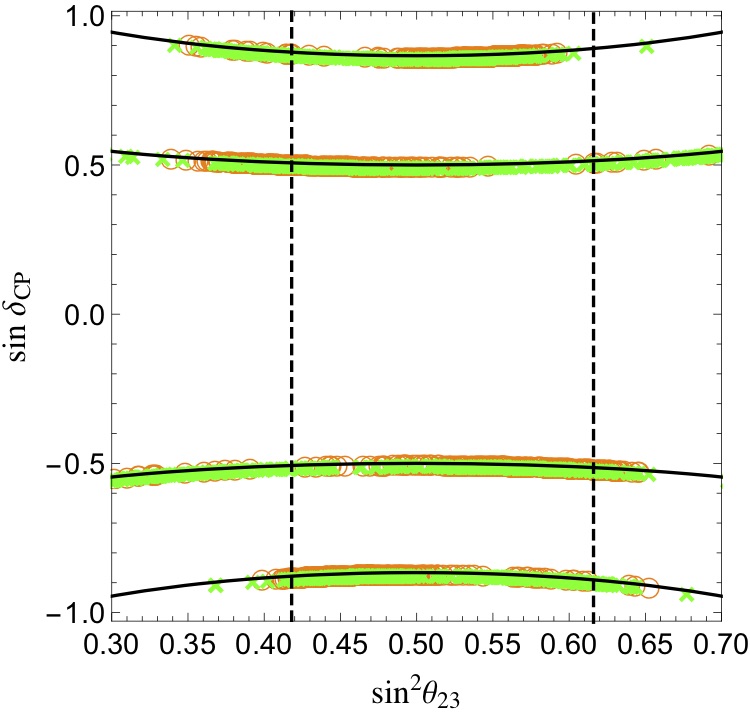}}\hspace*{0.1cm}
\subfigure{\includegraphics[width=0.48\textwidth]{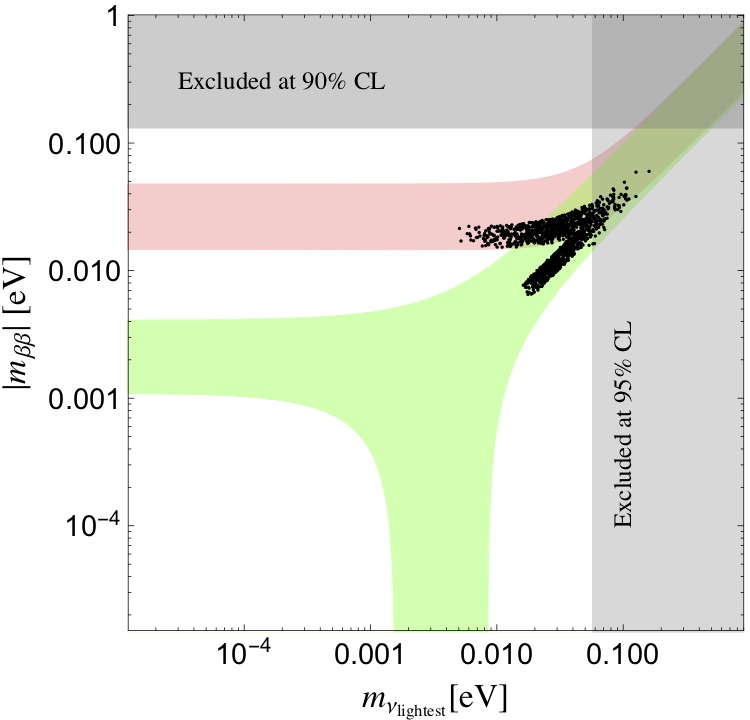}}
\caption{Same as Fig. \ref{fig3} but for $n=6$ (with $-\eta_1=\eta_2=1$, $\gamma =1$ and $x=1,2$).}
\label{fig4}
\end{figure}

The perturbations to $M_{0\nu}$ characterized by Eq.  (\ref{delmnu}) are also expected to induce viable solar neutrino mass scale. We investigate this by performing numerical study of the neutrino mass matrix $M_\nu$ given in Eq. (\ref{explicitmnu}). The parameters $A$, $B$ and $C$ are chosen from  uniform random distributions of real numbers in the range from $\pm 0.1$ to $\pm 1.0$. The perturbation parameters are also randomly chosen such that $|\epsilon_i| \le \epsilon_{\rm max}$ where $i=1,2,3$ and we study cases for $\epsilon_{\rm max} = 0.02$ or $0.1$. With this choice, it is ensured that the magnitude of perturbation parameters is always smaller than that of $A$, $B$ and $C$. Several sample points are generated and for each point the parameter $m_0$ in Eq. (\ref{explicitmnu}) is determined using the known value of atmospheric squared mass difference,  $\Delta m^2_{31} = 2.494 \times 10^{-3}\, {\rm eV}^2$  ($\Delta m^2_{23}= 2.465 \times 10^{-3}\, {\rm eV}^2$) in case of normal (inverted) ordering in 
the neutrino masses. Here, $\Delta m^2_{ij} \equiv m_i^2-m_j^2$ and we use the observed values from the latest global fit results NuFIT 3.2 (2018) \cite{Esteban:2016qun}. We then compute neutrino masses and mixing using numerical $M_\nu$ and $U_l$ given in Eq. (\ref{ul}). The results obtained for some small groups in $\Delta(6n^2)$ group series corresponding to $n=2,4,6$ are displayed in Figs. \ref{fig2}-\ref{fig4}.

The correlations between the effective mass of neutrinoless double beta
decay $|m_{\beta \beta}|$ and the lightest neutrino mass are displayed in
Fig. \ref{fig2} in case of $S_4$ group and for small and large 
perturbations. The region shown by the horizontal gray band is disfavoured
by the combined limit obtained from different neutrinoless double beta decay
experiments at $90 \%$ confidence level \cite{Guzowski:2015saa}. The
vertical gray band represents the region excluded by the limit on the sum of
neutrino masses obtained from the latest results from PLANCK experiment,
baryon acoustic oscillations and type Ia supernovae \cite{Couchot:2017pvz}.
The regions shaded by green and red colours are generically allowed when the
Dirac and Majorana phases are unconstrained in case of normal and inverted
ordering in neutrino masses respectively. The scattered points in Fig.
\ref{fig2} are in agreement with the $3\sigma$ ranges of global fit \cite{Esteban:2016qun}
values of solar and atmospheric mass differences and all three mixing 
angles. It can be seen that the small values of perturbations, corresponding to $\epsilon_{\rm max} = 0.02$, require nearly degenerate mass spectrum for all the three neutrinos which is almost disfavoured by the considered limit on the sum of neutrino masses. Increasing the magnitude of perturbations helps in evading this constraint. However one obtains lower limit on the lightest neutrino mass: $m_{\nu_1} > 0.015$ eV for normal ordering and $m_{\nu_3} > 0.005$ eV for inverted ordering if the magnitude of perturbation is smaller than that of the leading order parameters. This case predicts maximal values for $\theta_{23}$ and $\delta_{\rm CP}$ and vanishing Majorana phases irrespective of the magnitude of perturbation and therefore it is easily falsifiable in many ways.

We also perform similar numerical investigations for some cases which provide alternative predictions for $\theta_{23}$ and $\delta_{\rm CP}$. We choose the groups in $\Delta(6n^2)$ series corresponding to $n=4,6$. As it can be seen from Fig. \ref{fig1}, this choice offer three distinct values for $p$, one for $n=4$ and two for $n=6$, which are neither real nor imaginary. In these cases the predicted value of $\theta_{23}$ depends on the perturbations and the values of CP phases are determined by the correlation predicted in the first two of Eq. (\ref{correlations}). The results are displayed in Fig. \ref{fig3} for $n=4$ and in Fig. \ref{fig4} for $n=6$. 
It is noticed that correlation between $|m_{\beta \beta}|$ and the lightest neutrino mass remains almost similar in all these cases.

 There exists a lower limit on the lightest neutrino mass if the size of perturbations is restricted. As it can be seen from Eq. (\ref{mnureal}), this is a generic feature of obtained neutrino mass matrix independent of a choice of group parameters. From Eq. (\ref{mnureal}), one obtains the following expressions for the neutrino masses at the leading order in $\epsilon_i$:
\beqa \label{analytical}
m^2_{\nu_1} & = & m_0^2 \left( A^2 + B^2 + \epsilon_3 A^2 \left(1 - \sqrt{1 + \frac{B^2}{A^2}}\right) + {\cal O}(\epsilon_i^2)\right) \,, \nonumber \\
m^2_{\nu_2} & = & m_0^2 \left( A^2 + B^2 + \epsilon_3 A^2 \left(1 + \sqrt{1 + \frac{B^2}{A^2}}\right) + {\cal O}(\epsilon_i^2)\right) \,, \nonumber \\
m^2_{\nu_3} & = & m_0^2\, \left( C^2 + {\cal O}(\epsilon_i^2) \right)\,.
\eeqa
Note that splitting between $m_{\nu_1}$ and  $m_{\nu_2}$ is induced by only $\epsilon_3$ at the leading order while $\epsilon_{1,2}$ contribute at the second order. The above masses imply the following relation between the solar squared mass difference and $m_{\nu_1}$.
\be \label{analytical_msol}
\Delta m^2_{21} = m_{\nu_1}^2\, \left( \frac{2 \epsilon_3}{\sqrt{1+B^2/A^2}} + {\cal O}(\epsilon_i^2)\right)\,.\ee
The above relation implies that $m_{\nu_1}$ cannot be arbitrary small for finite value of $\epsilon_3$ for nonzero $\Delta m^2_{21} $. For ${\cal O}(1)$ values of parameters $A$ and $B$, one obtains $m_{\nu_1} \gsim 0.05$ ($0.015$) eV for $\epsilon_3 \lsim 0.02$ (0.1) from Eq. (\ref{analytical_msol}) which sets lower bound on the mass of the lightest neutrino in case of normal ordering. In case of inverted ordering, a lower bound on the lightest neutrino mass arises if $|C| > |\epsilon_i|$ is assumed as seen from Eq. (\ref{analytical}). The above observations are in agreement with the numerical results discussed earlier in this section.

\section{Summary}
\label{summary}
Flavour symmetries with or without CP have been widely used for prediction of the leptonic mixing angles and phases. We have shown here that the flavour dependent CP symmetries can also play a role in restricting neutrino masses and can lead to a partially or fully  degenerate neutrino spectrum. We have studied this in a general set up independent of any specific groups and discussed underlying constraints. The degeneracy follows from the conventional assumptions, Eqs. (\ref{sinv},\ref{xinv}) made in the standard approaches \cite{Feruglio:2012cw,Holthausen:2012dk,King:2014rwa} proposed to constrain neutrino mixing parameters. The only specific additional requirement is either Eq. (\ref{xs1s2}) or Eq. (\ref{xs1s2s3}) which imply degeneracy in the masses of two or three neutrinos respectively. We have elaborately discussed a case which leads to degenerate solar neutrino pair. The $\Delta(6 n^2)$ groups with even $n$ provide concrete examples of general set up discussed here. GenCP  symmetries  implying 
partially degenerate neutrinos within these groups were already discussed in \cite{King:2014rwa} but the occurrence of two degenerate neutrinos within them was not noticed. 

The symmetries envisaged here can also  be used for predictions of  the
neutrino mixing angles and phases. However, a complete determination of these
requires perturbations to break degeneracy in masses. We have discussed
specific perturbations in $\Delta(6 n^2)$  which break the residual Klein
symmetry of $M_\nu$ but preserve the underlying CP symmetry. Even after perturbations,
the GenCP invariance of the full mass matrix $M_\nu$ leads to predictions of
CP violating phases $\delta_{CP}$, $\alpha_{21}$, $\alpha_{31}$ in terms of a
group theoretical phase factor and the atmospheric mixing angle
$\theta_{23}$. The phenomenologically interesting scenario of residual generalized $\mu$-$\tau$ symmetry, which predicts maximal $\sin 2\theta_{23}$ and $\delta_{CP}$, is obtained as a special case here. Detailed predictions of  $\Delta(6 n^2)$ groups for $n=2,4,6$ are numerically studied taking into account constraints on neutrino masses and mixing parameters from the latest global fits to neutrino oscillation data. The smallest group $S_4$ predicts $\sin^2\theta_{23} = 1/2$ and $\delta_{CP}=\pm \pi/2$, even in the presence of CP invariant perturbations. The same predictions can also be obtained by any of the groups in $\Delta(6 n^2)$ series with even $n$. The groups with higher order can lead to progressively smaller values of $\sin\delta_{CP}$ as well. For all the cases studied here, it is found that the smallness of the size of perturbations puts lower bound on the mass of the lightest neutrino. Therefore, the predictions made here can be tested not only from the precise measurements of atmospheric angle 
and Dirac CP phase but also from the experiments sensitive to the absolute scale of the neutrino masses.

\section*{Acknowledgements}
This work was supported by BRNS (Department of Atomic Energy) and by
Department of Science and Technology, Government of India through the Raja Ramanna  fellowship and the J. C. Bose grant respectively. The work of KMP was partially supported by SERB Early Career Research Award (ECR/2017/000353) and by a research grant under INSPIRE Faculty Award (DST/INSPIRE /04/2015/000508) from the Department of Science and Technology, Government of India.

\bibliography{references}
\bibliographystyle{apsrev4-1}
\end{document}